%% file: elsarticle-template-num.tex
\documentclass[preprint,12pt]{elsarticle}

\usepackage{amssymb}
\usepackage{amsmath}
\usepackage[utf8]{inputenc}
\usepackage{amsmath,amsfonts}
\usepackage[ruled]{algorithm2e}

\usepackage{float}
\usepackage{array}
\usepackage{pbox}
\newcolumntype{?}{!{\vrule width 2pt}}
\usepackage{textcomp}
\usepackage{stfloats}
\usepackage{url}
\usepackage{verbatim}
\usepackage{graphicx}
\usepackage{multirow}
\usepackage{siunitx}
\usepackage{listings} 
\usepackage{physics}
\usepackage{graphicx}
\usepackage{upgreek}
\usepackage{adjustbox}
\usepackage{minted}
\usepackage{csquotes}
\usepackage{caption}
\usepackage{subcaption}

\usepackage{xcolor}

\RequirePackage{doi}
\usepackage{hyperref}
\usepackage[capitalise, nameinlink, noabbrev]{cleveref} 
\usepackage{qcircuit}
\usepackage{tikz}
\usepackage[framemethod=TikZ]{mdframed}
\usetikzlibrary{tikzmark}
\usetikzlibrary{tikzmark,decorations.pathreplacing}
\usetikzlibrary{backgrounds}
\usepackage{booktabs}
\newlength\mylen
\newcolumntype{C}{>{\hfil$}p{\mylen}<{$\hfil}}
\newcolumntype{F}[1]{>{\centering\arraybackslash}m{#1}} 

\usepackage{url}

\usepackage{breakurl}
\DeclareMathOperator\sgn{sgn}

\journal{Journal of Systems Architecture}

\begin{document}

\begin{frontmatter}



\title{High-Parallel FPGA-Based Discrete Simulated Bifurcation for Large-Scale Optimization}


\author{Fabrizio Orlando} 

\affiliation{organization={Politecnico di Torino},
            addressline={corso duca degli Abruzzi 24}, 
            city={Turin},
            postcode={10129}, 
            state={Italy},
            country={Italy}}

\author{Deborah Volpe} 

\affiliation{organization={Istituto Nazionale di Geofisica e Vulcanologia},
            city={Rome},
            state={Italy},
            country={Italy}}

\author{Giacomo Orlandi} 

\affiliation{organization={Politecnico di Torino},
            addressline={corso duca degli Abruzzi 24}, 
            city={Turin},
            postcode={10129}, 
            state={Italy},
            country={Italy}}

\author{Mariagrazia Graziano} 

\affiliation{organization={Politecnico di Torino},
            addressline={corso duca degli Abruzzi 24}, 
            city={Turin},
            postcode={10129}, 
            state={Italy},
            country={Italy}}

\author{Fabrizio Riente} 

\affiliation{organization={Politecnico di Torino},
            addressline={corso duca degli Abruzzi 24}, 
            city={Turin},
            postcode={10129}, 
            state={Italy},
            country={Italy}}

\author{Marco Vacca} 

\affiliation{organization={Politecnico di Torino},
            addressline={corso duca degli Abruzzi 24}, 
            city={Turin},
            postcode={10129}, 
            state={Italy},
            country={Italy}}

\begin{abstract}
Combinatorial Optimization (CO) problems exhibit exponential complexity, making their resolution challenging. Simulated Adiabatic Bifurcation (aSB) is a quantum-inspired algorithm to obtain approximate solutions to large-scale CO problems written in the Ising form. It explores the solution space by emulating the adiabatic evolution of a network of Kerr-nonlinear parametric oscillators (KPOs), where each oscillator represents a variable in the problem. The optimal solution corresponds to the ground state of this system. A key advantage of this approach is the possibility of updating multiple variables simultaneously, making it particularly suited for hardware implementation. To enhance solution quality and convergence speed, variations of the algorithm have been proposed in the literature, including ballistic (bSB), discrete (dSB), and thermal (HbSB) versions.\\
In this work, we have comprehensively analyzed dSB, bSB, and HbSB using dedicated software models, evaluating the feasibility of using a fixed-point representation for hardware implementation. We then present an open-source hardware architecture implementing the dSB algorithm for Field-Programmable Gate Arrays (FPGAs). The design allows users to adjust the degree of algorithmic parallelization based on their specific requirements. A proof-of-concept implementation that solves 256-variable problems was achieved on an AMD Kria KV260 SoM, a low-tier FPGA, validated using well-known max-cut and knapsack problems.
\end{abstract}







\begin{keyword}
Ising Machines \sep Simulated Bifurcation \sep FPGA \sep Parallel Computing \sep Combinatorial Optimization
\end{keyword}

\end{frontmatter}

\input{article_text}
\bibliography{acmart.bib}

\end{document}

%% file: article_text.tex
\section{Introduction}
\textbf{Combinatorial optimization} \textbf{(CO)} aims to determine the input configuration minimizing or maximizing an objective function. These problems often emerge in various practical domains such as resource allocation \cite{Obata2024allocation}, logistics \cite{Tsuyumine2024routing}, finance~\cite{Hong2021portfolio}~\cite{Venturelli2019portfolio}, and many others, i.e., whenever the solution minimizing or maximizing some figures of merit must be identified among a discrete set of feasible ones.\\
The main challenge in CO problems is the computational complexity required for solving them since many of those belong to the NP-hard class, i.e., the solution space grows exponentially with the problem size.\\
The Ising model is a mathematical formulation, inherently NP-complete, for describing CO problems.
A new group of hardware accelerators, referred to as \textbf{Ising machines}, have been designed to tackle challenging optimization problems described according to the homonymous model. These solvers have been developed using various technologies, including optical oscillators, digital logic, and quantum hardware. 
Among them, the recently proposed \textbf{Simulated Bifurcation Machines} (\textbf{SBMs})\cite{goto2019combinatorial}, implementing the solution space exploration through the \textbf{Simulated Bifurcation} (\textbf{SB}) algorithm, stands out for its implementability on digital architectures. This algorithm emulates the evolution of a network of Kerr non-linear parametric oscillators (KPOs), which exhibit bifurcation phenomena. The two branches of the bifurcation can be associated with the two states of a discrete variable. Initially, an adiabatic evolution of the system has been considered (aSB), while recently \textbf{ballistic} (\textbf{bSB}) and \textbf{discrete} (\textbf{dSB}) evolutions of the algorithm have been proposed \cite{goto2021high} to prevent analogue errors, which can affect the solution quality.\\
This article presents an \textbf{open-source} hardware architecture, described in SystemVerilog, that implements the \textbf{dSB} algorithm, potentially assisted by the heating mechanism, for low-tier \textbf{Field-Programmable Gate Arrays} (\textbf{FPGAs}) and adaptable for future \textbf{Application-Specific Integrated Circuit} (\textbf{ASIC}) implementations. To the best of our knowledge, this is the first open-source architecture of dSB. It is designed to be flexible, allowing users to define the algorithm's degree of parallelization according to their needs, and it can solve any Ising problem, unlike other SBMs that are limited to max-cut-like problems. The optimal number representation and algorithm parameters have been analyzed using software models written in C++. A proof-of-concept hardware implementation solving 256-variable problems has been demonstrated on a \textbf{AMD Kria KV260} SoM FPGA and validated using the well-known max-cut and knapsack problems.\\
The paper is organized as follows. Section \ref{sec:theory} presents the Ising model and explores the SB algorithm and its variants. The idea behind the proposed high-parallel architecture for FPGAs is introduced in Section \ref{sec:Motivations}. Section \ref{sec:Implementation} delves into the implementation details. Finally, Section \ref{sec:Results} presents the attained results and in Section \ref{sec:Conclusions} conclusions are drawn.
\section{Background/Theoretical foundations} \label{sec:theory}
This section introduces the Ising formulation along with two problem benchmarks, followed by a discussion of the Simulated Bifurcation algorithm and its variations.
\subsection{Ising model}
The \textbf{Ising} model \cite{naito4study} is a physical-mathematical model used to represent magnetism in matter. It describes a system of interacting magnetic \textbf{spins} ($s_i$) arranged in a lattice, where each spin can assume one of two discrete states based on its orientation: \textbf{+1} (\textbf{spin-up}) or \textbf{-1} (\textbf{spin-down}). The following Hamiltonian describes the energy of this system:
 \begin{equation}
    H(\mathbf{s}) = -\frac{1}{2}\sum_{i=1}^N \sum_{j=1}^N \textbf{J}_{ij}s_is_j - B\sum_{i=1}^N \textbf{h}_i s_i \, 
    \label{eq:Ising_Hamiltonian}
\end{equation}
where $s_i$ is the $i^\textrm{th}$ magnetic spin, $\textbf{J}$ is a symmetric matrix representing interactions among spins and $\textbf{h}$ is a vector describing the preferred orientation of a spin (up or down) with respect to the external magnetic field $B$.\\
Recently, Ising formulation has been extensively leveraged for embedding CO problems since evolving the system for reaching the ground state corresponds to looking for the problem's optimal solution.\\
This model is perfectly equivalent to the popular \textbf{Quadratic Unconstrained Binary Optimization} (\textbf{QUBO}) formulation \cite{glover2018tutorial}. The main difference is that the first involves bipolar binary variables, while the second involves unipolar ones, making the translation from one to the other possible by using the relation $s{i} = 2q_i-1$, where $q_{i}$ is the QUBO unipolar binary variable.\\
In this article, we consider two well-known problems, the max-cut and the knapsack, as benchmarks, whose Ising formulation is discussed below. 
\subsubsection{Max-cut}
It aims to \textbf{partition an undirect graph into two complementary subsets, $S$ and $\overline{S}$, maximizing the cut}, i.e., the \textbf{sum of edges joining the two sets}. Associating a spin variable $s_i$ for each node assuming +1 if it belongs to $S$ and -1 otherwise, the size of the cut is equal to: 
\begin{equation}
    C(s) = \sum_{i=0}^{N-1} \sum_{j=0}^{N-1} w_{ij} \frac{1-s_is_j}{2} \, ,
\end{equation}
where $N$ is the number of nodes in the graph and $w_{ij}$ the weight of the edge joining the $i^{\textrm{th}}$ and $j^{\textrm{th}}$ nodes. Consequently, the function to minimize is: 
\begin{equation}
    H(s) = \sum_{i=0}^{N-1} \sum_{j=0}^{N-1} w_{ij}s_is_j \, .
\end{equation}
In this work, the \href{https://web.stanford.edu/~yyye/yyye/Gset/}{G-Set} set is considered for benchmarking.
\subsubsection{Knapsack}
Its target is to define the best subset of objects belonging to a set of $N$ items, where each is characterized by a preference parameter $p_i$ and a weight $w_i$, to insert in a knapsack, maximizing the preference score without exceeding a weight threshold $\mathcal{W}$ \cite{knapsack}.  Associating a spin variable $s_i$ to each object, assuming value 1 if the object is in the set and -1 otherwise, the problem can be described as: 
\begin{align*}
    \text{maximize}   \ \sum_{i=1}^{N} c_i\frac{1+s_i}{2} \, ,\quad  \text{subject to} \ \sum_{i=1}^{N} w_i\frac{1+s_i}{2} \leq \mathcal{W}\, .
\end{align*}
Differently from the max-cut problem, the knapsack is constrained and demands both the $J$ matrix and $h$ vector components of the Ising formulation for its description. Moreover, for rewriting the inequality constraint in a penalty function form, it is required to transform it into an equality one by introducing auxiliary variables. Consequently, the problem Hamiltonian can be written as \cite{Lucas_2014}: 
\begin{equation}
   H=H_\textrm{cost}+\lambda H_\textrm{constraint} \, ,
\end{equation}
where $H_{cost}$ describe the maximization of the preference score and can be written as:
\begin{equation}
    H_\textrm{cost} = -\sum_{i=1}^N c_i \frac{1+s_i}{2}\, .
    \label{eq:knapasack_unconstrained}
\end{equation}
while  $H_\textrm{constraint}$ allows penalization of the solutions not satisfying the weight constraint and can be written as: 
\begin{align}
    H_\textrm{constraint} = \left(\sum_{n=1}^{\textrm{log}_2(\mathcal{W})} 2^n\frac{1+y_n}{2} + \sum_{i=1}^N w_i \frac{1+s_i}{2} -\mathcal{W} \right)^2\, .
    \label{eq:knapsack_constraint_ham}
\end{align}
where $y_n$ is an auxiliary variable and $\lambda$ is the penalty weight.\\
In this work, the \href{http://artemisa.unicauca.edu.co/~johnyortega/instances_01_KP/}{0/1 Knapsack} set of problems is considered for benchmarking.
\subsection{Simulated Bifurcation}
 \begin{figure}[h]
    \centering
    \begin{subfigure}{0.3\textwidth}
        \centering
        \includegraphics[width=\textwidth]{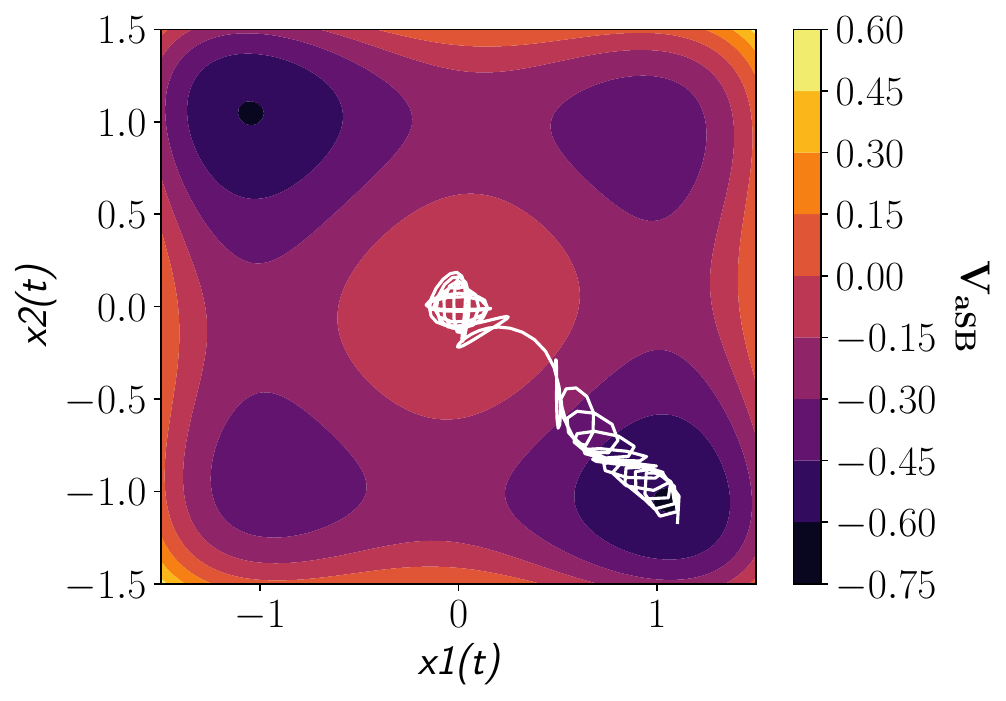}
        \caption{Simulated Adiabatic Bifurcation}
        \label{fig:adiabatic_traj}
    \end{subfigure} \quad
    \begin{subfigure}{0.3\textwidth}
        \centering
        \includegraphics[width=\textwidth]{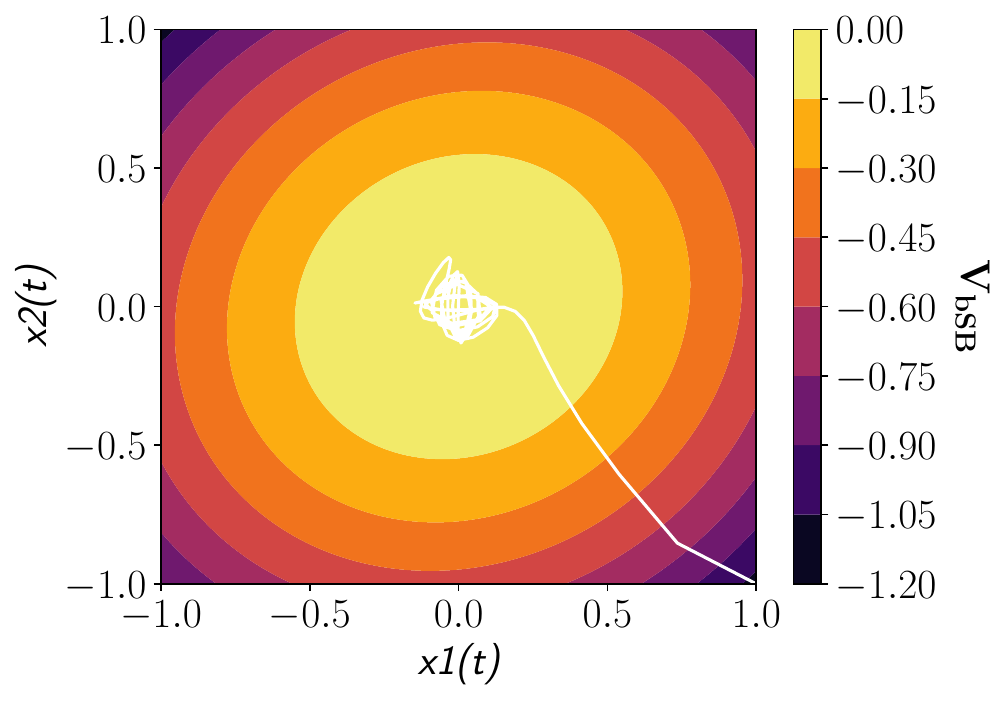}
        \caption{Ballistic Simulated Bifurcation}
        \label{fig:ballistic_traj}
    \end{subfigure} \quad
    \begin{subfigure}{0.3\textwidth}
        \centering
        \includegraphics[width=\textwidth]{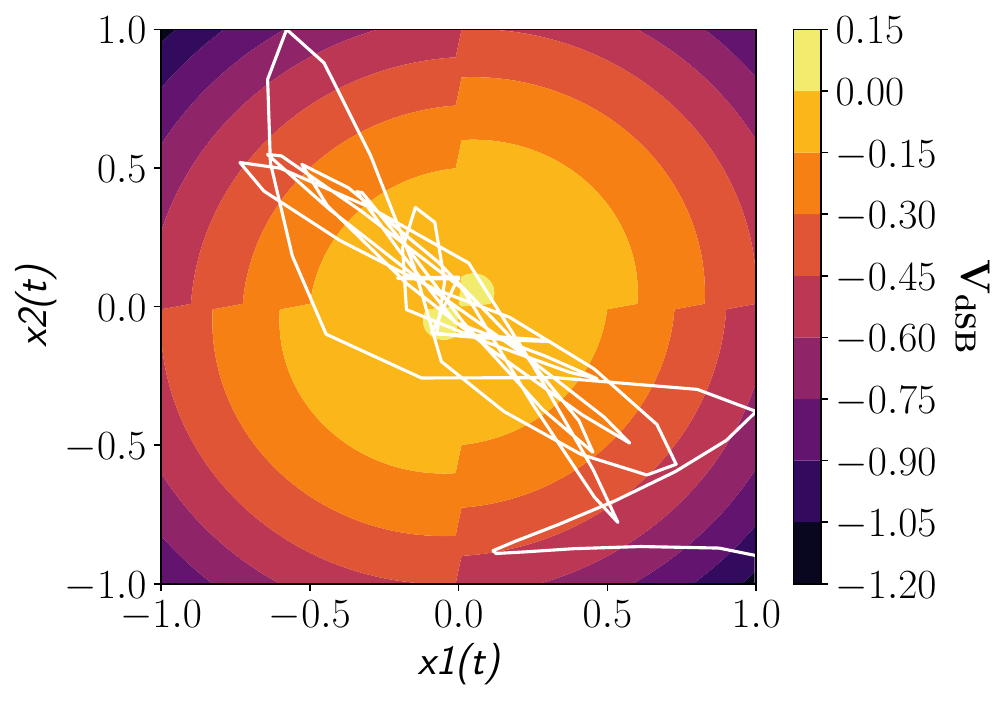}
        \caption{Discrete Simulated Bifurcation}
        \label{fig:discrete_traj}
    \end{subfigure}
    \label{fig:SB_traj}
    \caption{Algorithms comparison for a 2-node max-cut problem. The trajectories of the oscillators' network over time are shown by the white lines. The potential energies at the final time are also depicted ($V_{aSB}$, $V_{bSB}$ and $V_{dSB}$).}
\end{figure}
\begin{figure}[h]
    \centering
    \includegraphics[width=0.7\textwidth]{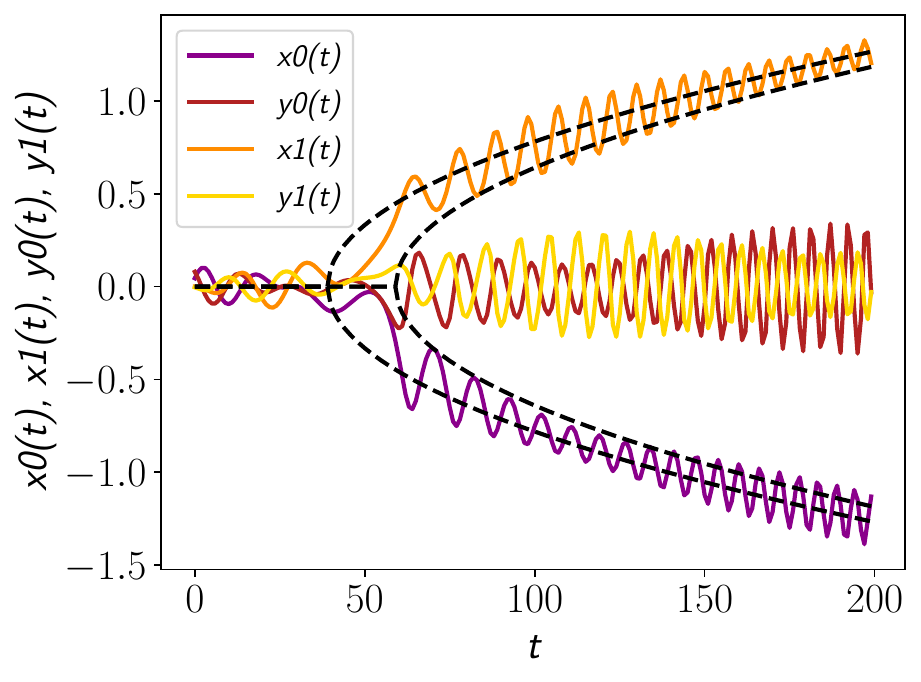}
    \caption{Position and momentum variable evolutions
of a two oscillators system with anti-ferromagnetic interaction.}
    \label{fig:aSBoscillators}
\end{figure}

\noindent \textbf{Simulated Adiabatic Bifurcation} (\textbf{aSB}), a quantum-inspired algorithm introduced in \cite{goto2019combinatorial}, offers approximate solutions of large-size optimization problems \textbf{in a limited amount of time} written according to Ising formulation. It mimics on classical platforms the quantum adiabatic evolution of a \textbf{non-linear-Kerr-oscillators network} excited by an input pumping signal ($a(t)$). These oscillators---each described by a pair of variables \textbf{position} ($x_i$) and \textbf{momentum} ($y_i$)---exhibit a bifurcation during their evolution obtained by gradually increasing $a(t)$ from zero to its final value ($a_0$), and each branch can be associated with a spin state \cite{goto2016bifurcation}. Therefore, a system of $n_{\textrm{spin}}$ oscillators can represent $2^{n_{\textrm{spin}}}$ energy states.  The problem is encoded by associating an oscillator with each spin variable, and the spin interactions are expressed through the network,  whose role is to create an imbalance in the energy of the systems such that the optimum of the problem corresponds to the final ground state, forcing each oscillator to choose the branch representing the spins state of the problem solution (Fig. \ref{fig:adiabatic_traj} and \ref{fig:aSBoscillators}). 
The evolution of the physical system can be described as:
\begin{equation}
    \begin{aligned}
        H_{SB} = \sum_{i=1}^{n_{\textrm{spin}}} \frac{\Delta}{2}y_i^2 + \sum_{i=1}^{N} \left[\frac{K}{4}x_i^4 + \frac{\Delta-a(t)}{2}x_i^2\right]+ \\ - \frac{c_0}{2}\sum_{i=1}^{n_{\textrm{spin}}}\sum_{j=1}^{n_{\textrm{spin}}}J_{i,j}x_ix_j\, ,
    \end{aligned}
\end{equation}
where $\Delta$ is the difference between the resonance frequency of each oscillator, assumed to be the same for all of them, and half the pumping frequency (of $a(t)$), $c_0$ is a positive constant and $K$ is the positive Kerr coefficient. \\
Deriving the equation of motion, discretizing the time in time-steps $\Delta_t$, and applying Euler's method, the following equation  can be obtained for the update of position and momentum variables during the system evolution:
\begin{equation}
x_i(t_{n+1}) = x_i(t_n) + \Delta y_i (t_n) \Delta_t\, ,
\label{eq:PositionUpdate}
\end{equation}
\begin{equation}
    \begin{aligned}
        y_i(t_{n+1}) =  y_i(t_n) - [Kx_i^3(t_{n+1}) + (\Delta -a(t_{n+1}))x_i(t_{n+1}) + \\ c_0 \sum_{j=1}^{n_\textrm{spin}} J_{ij}x_j(t_{n+1})] \Delta_t\, ,
        \label{eq:MomentumUpdate}
    \end{aligned}
\end{equation}
where $t_n$ is the $n^{\textrm{th}}$-time-instant ($t_n = n \Delta_t$).\\
A key advantage of this approach is the high level of parallelizability in simulating the system evolution, which promotes its hardware implementations such as on FPGA or ASIC. More details about the algorithm are available in \cite{volpe2024improving}. \\
However, the mathematical model employed for emulating the adiabatic evolution of the system generates some analogue errors,  potentially compromising performance. In response,  alternative approaches like the ballistic (bSB) and discrete (dSB) evolution of the network were introduced in \cite{goto2021high}.  \\
The bSB introduces for each oscillator $i$ two perfectly inelastic walls at $x_i = \pm 1$, as shown Fig. \ref{fig:ballistic_traj}. These walls are implemented by setting $x_i = \textrm{sgn} (x_i)$ and $y_i = 0$, whenever $|x_i| > 1$ in the equations for the evolution of the state variables (\ref{eq:PositionUpdate}, \ref{eq:MomentumUpdate}). This way, the position variable is forced to assume a discrete value when the pumping signal increases, reducing the analogue errors. 
The inelastic wall plays the role of the potential wall in the aSB, allowing the removal of the fourth-order term in $H_{SB}$. \\
The dSB has been proposed to further remove the analogue errors, discretizing the bSB. In particular, the singularity on the boundaries between positive and negative regions has been intentionally neglected, violating energy conservation across boundaries and escaping from local minima over potential barriers, as shown in Fig. \ref{fig:discrete_traj}. This is implemented substituting $x_j(t_{n+1})$ with $\sgn{(x_j(t_{n+1}))}$ in $\sum_{j=1}^{n_{\textrm{spin}}} J_{ij}x_j(t_{n+1})$ component of Eq. \ref{eq:MomentumUpdate}.\\
Both bSB and dSB preserve aSB's parallizability advantage, improving speed convergence and accuracy at the same time.\\
The bSB can be further enhanced by introducing a positive thermal fluctuation term called $\gamma$ (HbSB) for escaping from local minima, as proposed in \cite{kanao2022simulated}. The term is considered as an additional component $\gamma y_i(t_n)$ in Eq. \ref{eq:MomentumUpdate}.\\
The following compact equations can describe the evolution of the system of oscillators for all the algorithm variants:
\begin{align}
    \begin{cases}
	\tilde{y}_i & = y_i(t_n) + \{-[a_0-a(t_k)]x_i(t_k) +                         c_0f_i\}\Delta t\, ,\\
	\tilde{x}_i & = x_i(t_n) + a_0\tilde{y}_i\Delta t\, ,\\
	f_i & = \begin{cases}
	    \sum_{j=1}^{n_\textrm{spin}} J_{i,j}x_j & \text{for bSB}\, ,\\
            \sum_{j=1}^{n_\textrm{spin}} J_{i,j}\sgn(x_j) & \text{for dSB}\, ,
	\end{cases}
    \end{cases}
    \label{eq:discrete_pos_mom}
\end{align}
\begin{subequations}
    \begin{align}
        \label{eq:x_wall_discrete}
        x_i(t_{n+1}) &= 
        \begin{cases}
           \tilde{x}_i & \text{if } |\tilde{x}_i| \leq 1\, , \\
           \sgn(\tilde{x}_i) & \text{otherwise}
        \end{cases}\\
        y_i(t_{n+1})  &= 
        \begin{cases}
           \tilde{y}_i + \gamma y_i(t_k)\Delta t & \text{if } |\tilde{x}_i| \leq 1  \, , \\
           \gamma y_i(t_k)\Delta t & \text{otherwise}
        \end{cases}
    \end{align}
\end{subequations}
All the mentioned variants of the algorithm have an execution time for a sequential implementation that scales with the number of spins as:
\begin{equation}
    t_{\textrm{exec}} = \mathcal{O}(n_\textrm{steps}\cdot(n_{\textrm{spin}}^2 + n_{\textrm{spin}})\cdot T_\textrm{ck})\, ,
    \label{eq:sequential}
\end{equation}
where $n_\textrm{steps}$ is the number of algorithm steps necessary for reaching convergence and $T_\textrm{ck}$ the clock period.\\
Some hardware implementations of the Simulated Bifurcation algorithm, called \textbf{Simulated Bifurcation Machines} (\textbf{SBMs}), have already been designed and presented in the state of the art. An FPGA implementation of the aSB algorithm is introduced in \cite{tatsumura2019fpga} and further scaled using a multi-chip architecture in \cite{tatsumura2021scaling}. Additionally, two other architectures are discussed in the literature: \cite{volpe2024improving}, which, to the best of our knowledge, is the only open-source architecture currently available for the aSB algorithm, and \cite{zou2020massively}, which is specifically optimized for \textit{sparse} Ising problems.
\section{Towards a High-Parallel Ising Machine} \label{sec:Motivations}
This section discusses the motivations behind this work and the challenges and unmet needs it aims to address.
\subsection{Motivations}
Solving Ising problems effectively requires fast and accurate Ising machines. SBMs offer potential through massive parallelization, but the lack of open-source implementations, especially for ballistic, discrete, and heated variants, limits their evaluation in optimization contexts. Additionally, most studies focus on the max-cut problem, which only uses the \textbf{J} matrix and does not fully reflect broader optimization challenges. To fully assess the algorithm's potential, integrating the \textbf{h} vector and testing it on diverse benchmarks is essential. In fact, achieving a balance between speed, accuracy, and efficiency is key to ensuring scalability and high-quality results.
\subsection{General Idea}
\noindent This work aims to provide a versatile and comprehensive open-source SBM architecture. Its generic design, encompassing number representation and degrees of parallelization, enables synthesis on FPGAs for on-premises solutions. Software models, available on the GitHub \cite{gitHubRepo} repository with the hardware description, have been used to compare algorithm variations, select the optimal compromise for hardware implementation, study the impact of number representation, and analyze correlations among algorithm parameters. Additionally, an approach has been developed to incorporate the $\textbf{h}$ vector in the optimization process. The proposed open-source software implementation and architecture enable users to evaluate the potential of SBMs for any type of Ising problem.

\section{Implementation} \label{sec:Implementation}
This section presents the software analysis, algorithm parallelization, and design choices of the proposed architecture.
\subsection{Software analysis}
Starting from the algorithm description of Eq. \ref{eq:discrete_pos_mom} and the pseudocode \ref{alg:sequential}, a \texttt{C++} model of the bSB, HbSB, and dSB algorithm have been obtained both considering floating and fixed-point number representation.\\
The model requires some parameters that significantly impact the outcomes. $\Delta_t$ and $a_0$ are externally defined, and the entire algorithm is iterated for $n_\textrm{steps}$ number of times, which determines the rate at which the pumping signal varies from $0$ to $a_0$. Therefore, a finer simulation step corresponds to a higher $n_\textrm{steps}$ yielding greater accuracy in the results. $c_0$ can be automatically defined to have the first bifurcation point close to $0$ to have the fastest possible convergence. As discussed in \cite{goto2019combinatorial}, this point corresponds to the eigenvector linked to the largest eigenvalue of the $J$ matrix, which can be approximated with the expression of $\lambda_{\textrm{MAX}}$. Hence, the equation for $c_0$ is the following:
\begin{equation}
    c_0 = \frac{\Delta}{\lambda_{\textrm{MAX}}}, \quad \lambda_{\textrm{MAX}} \approx 2\sigma \sqrt{n_\textrm{spin}}\, ,
\end{equation}
where $\Delta$ is related to the detuning frequency of the pumping signal (set to $1$ for the discrete and ballistic cases \cite{goto2021high}), and $\sigma$ is the standard deviation of the $J$ matrix elements.\\
Position and momentum variables must be initialized close but not equal to $0$ to stimulate the KPOs, emulating environmental noise. The algorithm is run multiple times to select the best results obtained from different initialization values.\\
To account for the \textbf{h} component of the problem, we adopt the solution proposed in \cite{zhang_date}, introducing an \textbf{ancillary variable} to rewrite the Hamiltonian as:
\begin{align}
H^*(s) &= -\frac{1}{2}\sum_{i=1}^{N}\sum_{j=1}^{N}J_{i,j}s_is_j - \sum_{i=1}^{N} h_is_i\cdot s_{N+1} \\&= -\frac{1}{2}\sum_{i=1}^{N}\sum_{j=1}^{N} J_{i,j}^{\star}s_is_j \, ,
\end{align}
which allows the integration of \textbf{h} into the \textbf{J} matrix as an additional row and column: \begin{equation*}
    \underline{\underline{J^{\star}}} = 
    \left[\begin{array}{ccccc|c}
    0 & J_{12}  & J_{13}& \ldots & J_{1N} &h_1 \\
    J_{12} & 0 & J_{23}&\ldots&J_{2N}  &h_2 \\
    J_{13} & J_{23} & 0 & \ldots & J_{3N} & h_3\\
    \vdots & \vdots & \vdots & \ddots& \vdots&\vdots \\
   J_{1N} & J_{2N} & J_{3N} &\ldots&0 &h_N \\\hline
    h_1 & h_2 & h_3 &\ldots&h_N &0 \,
\end{array}\right] \, .
\end{equation*}
\input{pseudocode}
\noindent To maintain equivalence with the original Ising model, $s_{N+1}$ must equal to 1, ensuring $x_{N+1}$ remains positive throughout the algorithm's evolution. For bSB, it is convenient to gradually increase the ancillary variable over time to balance the contributions of \textbf{J} and \textbf{h}, while for dSB, it can be directly fixed at 1, as it evaluates only the sign.
\begin{figure}[htpb]
    \centering
    \includegraphics[width=0.7\textwidth]{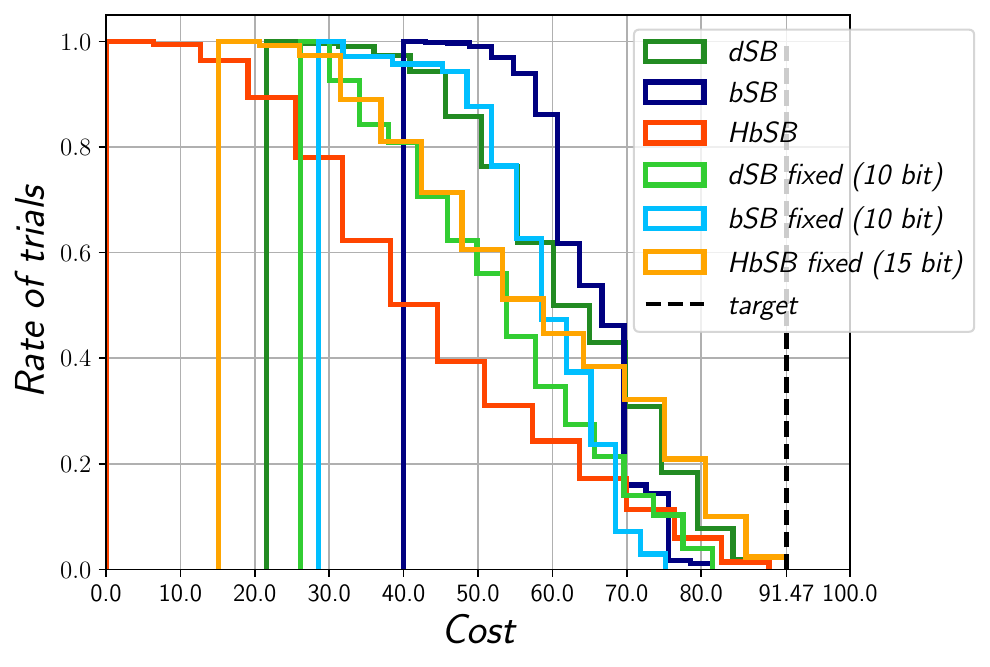}
    \caption{Cumulative distribution obtained for a \href{http://artemisa.unicauca.edu.co/~johnyortega/instances_01_KP/}{100 objects knapsack problem} considering fixed and floating point implementation of dSB, bSB and HbSB.}
    \label{fig:cumulative}
\end{figure}
At this stage, both fixed-point and floating-point models were developed to evaluate if the accuracy loss was minimal enough to take advantage of the speed and memory occupation benefits. For the fixed-point implementation, the number of bits required for the fractional part is determined by the $da$, the smallest number to be represented. On the contrary, the number of bits of the integer part depends on $f_i$, the maximum number to be stored. The difference in accuracy between using fixed-point numbers and floating point numbers is minimal, as shown in Fig. \ref{fig:cumulative}. Therefore, fixed-point representation has been selected for the hardware implementation.\\
\begin{figure}[htpb]
    \centering
    \includegraphics[width=0.7\textwidth]{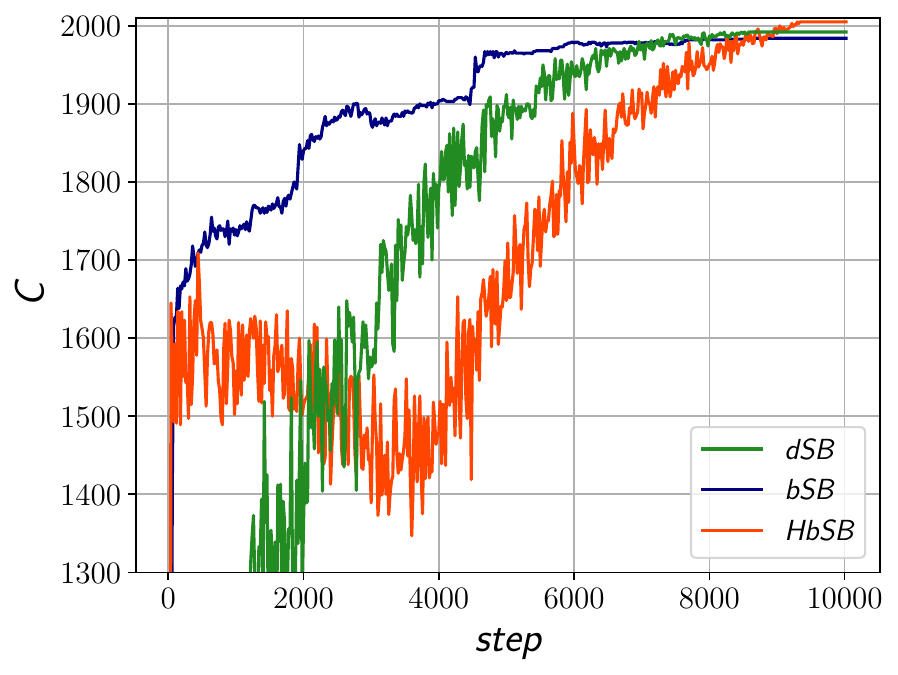}
    \caption{Comparison of the evolution of the cut value over time with the 800-spin G7 max-cut problem of the \href{https://web.stanford.edu/~yyye/yyye/Gset/}{GSet}, using $\Delta t=1.0$, $\gamma=0.5$ and $n_\textrm{steps} = 10^4$, for dSB, bSB and HbSB.}
    \label{fig:cost}
\end{figure}
Additionally, the model was used to assess the impact of parameter values on the different algorithm variations. Specifically, the choice for $\Delta_t$ depends on the algorithm variant and the problem addressed. As expected, increasing the number of steps, the results move closer to the target. For the max-cut problem of Fig. \ref{fig:cost}, $\Delta_t = 1.0$ was optimal for the dSB, whereas $\Delta_t = 0.5$ worked best for bSB. The ideal choice for $n_\textrm{steps}$ is a compromise between accuracy and speed, e.g., $n_\textrm{steps} = 10^4$ is a suitable choice for this case. Furthermore, increasing the fluctuation coefficient ($\gamma$) leads to greater fluctuations in the variables, which requires a higher number of steps to reach convergence.\\
Otherwise, the accuracy of the results can be degraded.\\
Finally, the selection of the most suitable algorithm variant for hardware implementation was made according to three figures of merit:
\begin{itemize}
    \item \textbf{Accuracy} (distance of the results distributions from the target value): The HbSB exhibits the best accuracy thanks to thermal fluctuations, the dSB achieves similar but slightly worse results, and the bSB has very tight distributions but the worst accuracy, as illustrated in Fig. \ref{fig:cumulative} and \ref{fig:cost}.
    \item \textbf{Speed} (average number of algorithm steps needed for convergence): the bSB is the fastest to reach a local minimum, followed by the dSB, while the HbSB is the slowest since it requires a higher number of steps for larger $\gamma$, as shown in Fig. \ref{fig:cost}.
    \item \textbf{Area} (number of hardware resources): in the dSB the value of the position variable is replaced with its sign, as a consequence it does not need multipliers to perform the matrix-vector multiplication. Hence, the saved resources can be employed to exploit parallelization, unrolling the algorithm since all the variables can be updated simultaneously.
\end{itemize}
Therefore, dSB has been chosen for hardware implementation as the best compromise in terms of speed, accuracy and the required area, since it is the most efficient for hardware resources requirement and very close to the best for accuracy and speed. Given the potential accuracy improvements observed using thermal fluctuations in bSB, and considering the minimal impact on hardware area, we decided to integrate a heating mechanism into the dSB hardware implementation. This addition, which has not been explored in the literature and was not included in the software models, could further enhance dSB's performance for some critical applications.
\subsection{Algorithm parallelization}
\label{sec:alg_parall}
\input{alg_parallelization/alg_parallelization}

\subsection{Architecture}
\input{alg_parallelization/matrix}
\begin{table}[H]
    \centering
    \caption{Execution time in terms of number of clock cycles per algorithm step using different parallelization parameters values and according with Eq. \ref{eq:time_pipeline}.}
    \label{tab:parameters}
    \begin{tabular}{F{1.1cm}|F{1.1cm}|F{1.1cm}|F{1.1cm}|F{1.7cm}}\toprule
    \textbf{$n_\textrm{spin}$} & \textbf{Pr} & \textbf{Pc} & \textbf{Pb} & \textbf{cycles/step}\\\midrule
    \multirow{3}{*}{256} & 64 & 4 & 4 & 80 \\
	& 8 & 16 & 4 & 132\\
	& 4 & 64 & 4 & 65\tikzmark{worse}\\
	& 16 & 16 & 4 & 68 \tikzmark{better} \\\bottomrule
    \end{tabular}
    \end{table}
    	\begin{tikzpicture}[overlay, remember picture]
		\draw[green, thick, rounded corners, fill=green, nearly transparent] ([shift={(1cm,2ex)}]pic cs:better)  rectangle ([shift={(-5.8cm,-.5ex)}]pic cs:better);
		\draw[red, thick, rounded corners] ([shift={(1cm,1.5ex)}]pic cs:worse)  rectangle ([shift={(-5.8cm,-.5ex)}]pic cs:worse);
	\end{tikzpicture}
A high-level description of the proposed architecture, inspired by that proposed in \cite{tatsumura2019fpga} for aSB implementation, is presented in Fig. \ref{fig:Architecture}. The main blocks are:
\begin{itemize}
    \item $\mathbf{P_b}$ MMTE units, each of them composed of a Matrix-vector Multiplication (MM) block in charge of computing $\sum_j J_{i,j}\cdot \sgn{x_j}$ and a Time Evolution (TE) block determining the update values of $x$ and $y$ variables;
    \item two memories (XMEM and YMEM) storing $x$ and $y$ values, supplied as inputs to the datapath (DP) along with the algorithm parameters ($\gamma$, $\Delta t$ and $c_0$);
    \item a linear updater increasing the pumping signal $a(t)$;
    \item two memory units (SGNXMEM1 and SGNXMEM2), storing the sign values of $x$ variables;
    \item memories storing the $J$ matrix coefficients. 
\end{itemize}
\input{alg_parallelization/alg_parallelization}
\subsubsection{Matrix-vector Multiplication unit}\hfill\\
\indent Each MM unit contains $\mathbf{P_r}$ Multiply ACcumulate (MAC) blocks, able to compute $\sum_j J_{i,j}\cdot \sgn{x_i}$. The scheme of a MAC unit is illustrated in Fig. \ref{fig:Architecture}. Each of them is fed by a $J$ memory, able to provide $\mathbf{P_c}$ elements of the $J$ matrix in parallel, and by another memory (SGNX) storing the sign values of the $x$ variables. The MAC unit performs $\mathbf{P_c}$ products between $J$ coefficients and the sign of $x$ values simultaneously. The partial products $J_{i,j}\cdot \sgn{x_j}$ can be computed using a multiplexer as in Fig. \ref{fig:Architecture}, where $\sgn{x_j}$ is used to select between $J_{i,j}$ and its 1's complement. These products are summed in a single step exploiting a multi-operand adder implemented as a binary tree of adders (Tree1). The 2's complement of the coefficients is eventually computed by adding the sum of the sign of $x$ values to the result using a second tree adder (Tree2) that works in parallel with the first one. Consequently, the result is accumulated (ACC) and sampled after $\frac{n_\textrm{spin}}{\mathbf{P_c}}$ clock cycles. The output MUX in Fig. \ref{fig:Architecture} enables to share the time evolution logic, performed by DP, among $\mathbf{P_r}$ paths.\\
\begin{figure}[b]
    \centering
    \includegraphics[width=\textwidth, trim=0cm 9cm 0cm 0cm,clip]{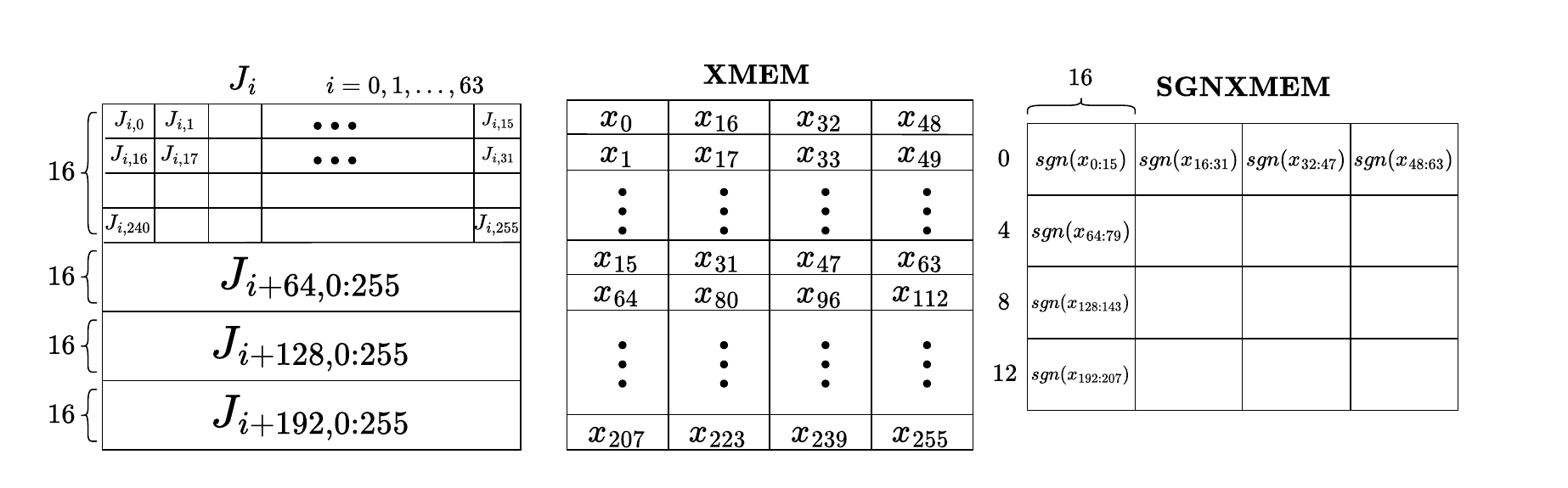}
    \caption{Organization of the three storing units inside the architecture when $Pr=Pc=16$ and $Pb=4$.}
    \label{fig:MEMS}
\end{figure}
\subsubsection{Memory organization}\hfill\\
\indent The architecture is composed of different storage units, as shown in Fig.~\ref{fig:MEMS}:
\begin{itemize}
    \item XMEM and YMEM have been implemented as simple dual RAMs with one port dedicated to reading and one to writing. Their data widths are $\mathbf{P_b} \cdot X_{\textrm{bits}}$ and $\mathbf{P_b} \cdot Y_{\textrm{bits}}$, respectively, to allow access to one $x$ and one $y$ variable for each MMTE unit at every clock cycle.
    \item The number of instantiated $J$ memories is $\mathbf{P_b} \cdot \mathbf{P_r}$, with each MAC unit having its own memory. These memories have a data width of $J_{\textrm{bits}} \cdot \mathbf{P_c}$, as they read $\mathbf{P_c}$ coefficients in parallel.
    \item SGNXMEM1(2) implemented as register files. Two of these are required: one stores $\sgn{x_i}$ at time $t_n$, while the other stores their updated values $\sgn{x_i(t_{n+1})}$.
\end{itemize}
\section{Results} \label{sec:Results}
\begin{figure}[h]
    \centering
    \includegraphics[width=0.7\textwidth]{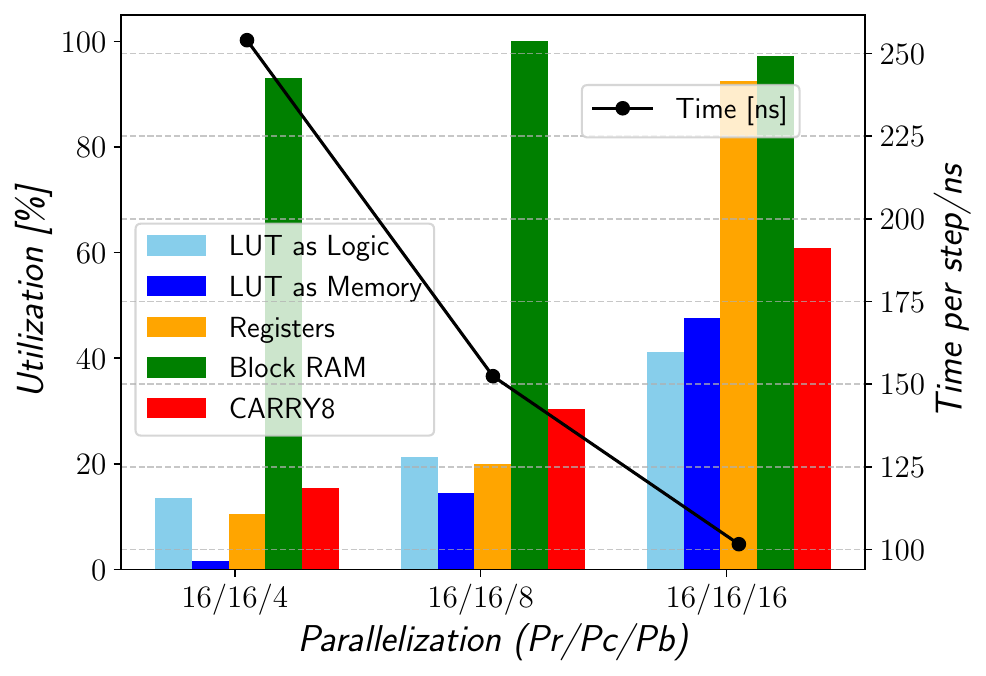}
    \caption{FPGA resource utilization for different parallelization parameters and time required to execute one algorithm step. The time on the left is obtained experimentally, while the other two are estimated values.}
    \label{fig:util}
\end{figure}

\begin{figure}[h]
    \centering
    \includegraphics[width=0.7\textwidth]{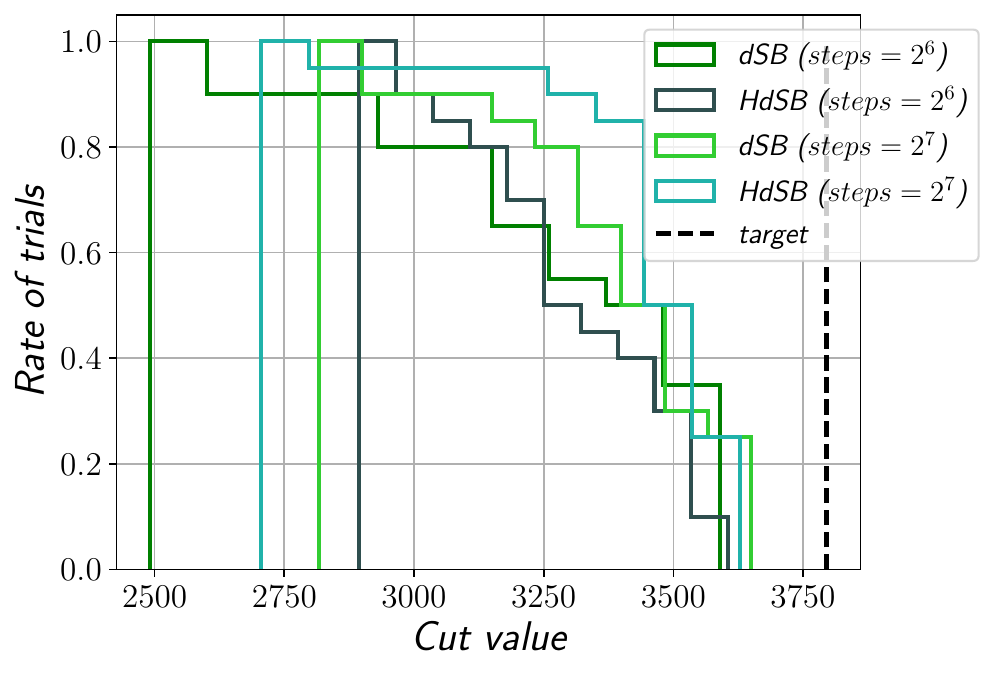}
    \caption{Results distribution for a randomly
generated 256-spin max-cut problem.}
    \label{fig:maxcuthardware}
\end{figure}

This section presents the synthesis results, optimization quality reached, and considerations on the parallelizability of the proposed architecture. The synthesis was performed on the \textbf{AMD Kria KV260 SoM} using \textbf{Vivado 2023.1} with default synthesis directives. Fig. \ref{fig:util} illustrates the FPGA resource utilization for different parallelization parameter choices. The proof-of-concept implementation was configured with ($\mathbf{P_r}=\mathbf{P_c}=16$ and $\mathbf{P_b}=4$). For each combination of parallelization parameters, the following FPGA resources were measured:
\begin{itemize}
    \item Look-Up Tables (LUTs) used as \textbf{logic blocks};
    \item LUTs used for \textbf{memory} units;
    \item Number of \textbf{registers};
    \item \textbf{CARRY8} blocks, which implement fast carry logic in arithmetic operations;
    \item \textbf{Block RAMs} (BRAMs).
\end{itemize}
The architecture's bottleneck is the available memory, as the allocated logic blocks account for less than 20\% of the available resources, while BRAM utilization exceeds 90\%, as shown in Table \ref{tab:fpga_utilization}. This highlights the fact that memory usage scales with the degree of parallelization. As parallelism increases, more J elements need to be accessed simultaneously, leading to a different organization of the coefficients in memory, thus requiring a greater number of the available memory blocks. The chosen configuration (shown on the left) represents the best trade-off within the available BRAM capacity. Fig. \ref{fig:util} also shows how the computation time scales as $\mathbf{P_b}$ increases.
\begin{table}[h]
    \centering
    \caption{FPGA Resource Utilization report with a maximum frequency of \SI{200}{\mega\hertz}, considering the parallelization parameters equal to $Pr=Pc=16$ and $Pb=4$ and the number of bits utilized to represent $J$ coefficients equal to 8.}
    \label{tab:fpga_utilization}
    \begin{tabular}{l|ccccl}
        \toprule
        \textbf{Block Type} & \textbf{Used} & \textbf{Available} & \textbf{Util\%} \\
        \midrule
        CLB LUTs                   & 16766 & 117120 & 14.32 \\
        \hspace{.8cm}LUT as Logic               & 15863 & 117120 & 13.54 \\
        \hspace{.8cm}LUT as Memory              & 903   & 57600  & 1.57  \\
        \hspace{1.6cm}LUT as Distributed RAM     & 626   & -      & -     \\
        \hspace{1.6cm}LUT as Shift Register      & 277   & -      & -     \\\midrule
        CLB Registers              & 24731 & 234240 & 10.56 \\
        \hspace{.8cm}Register as Flip Flop      & 24731 & 234240 & 10.56 \\\midrule
        CARRY8                     & 2259  & 14640  & 15.43 \\
        \midrule
        Block RAM Tile             & 134   & 144    & 93.06 \\
        \bottomrule
    \end{tabular}
\end{table}
\begin{figure}
    \centering
    \includegraphics[width=0.7\textwidth]{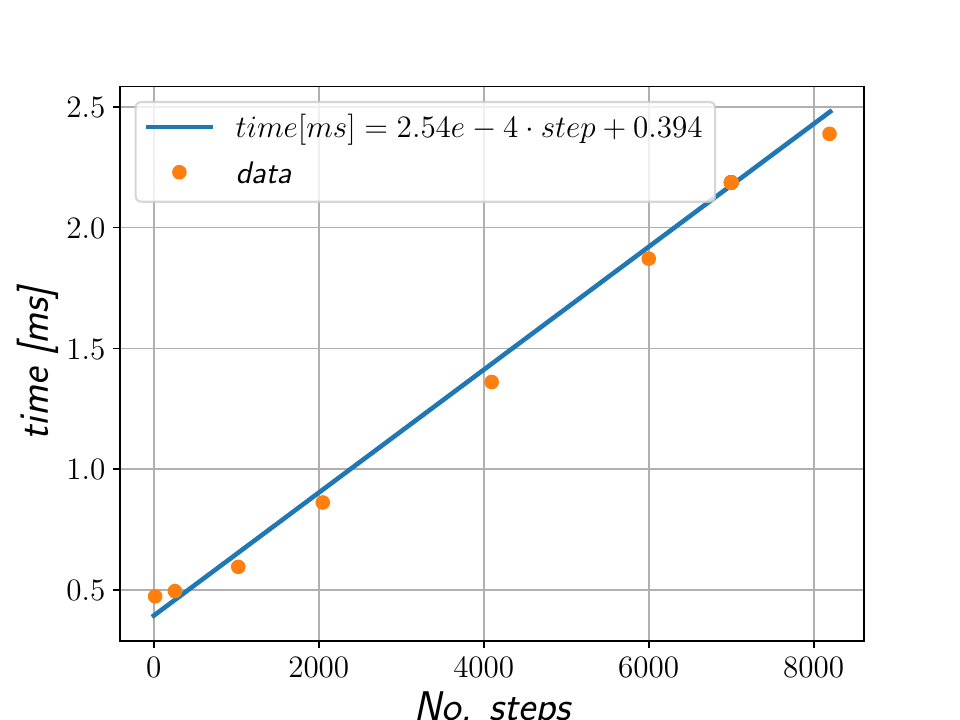}
    \caption{Time required to collect FPGA results versus number of algorithm steps. The orange dots represent the collected time data, whereas the blue curve is the best-fitting line obtained using the least squares method.}
    \label{fig:time}
\end{figure}
The timing data collected for the proof-of-concept implementation, as a function of algorithm steps, is shown in Fig.~\ref{fig:time}. The \textbf{time per step}, represented by the slope of the fitted line, is \SI{254}{\nano\second}. However, a significant \textbf{overhead} of approximately \SI{0.39}{\milli\second} is observed, primarily due to the initialization of all memory units prior to computation. When executing the algorithm multiple times on the same data, this initial overhead can be mitigated, as the J matrix only needs to be loaded once, though the position and momentum variables must still be reinitialized for each repetition.\\
The accuracy of the hardware implementation is shown in Fig. \ref{fig:maxcuthardware}, which illustrates the result distribution for a randomly generated 256-spin max-cut problem with $J$ coefficients ranging from -128 to 0. The target value, computed using a software model implementing Simulated Annealing (SA), is also included for comparison. As the number of steps increases, the distribution of results approaches the target. Notably, the algorithm achieves good approximations of the target in a relatively short time. For instance, 85\% of the \textit{cut values} obtained by running HdSB for 128 steps reach at least 90\% of the target value, with a computation time of \SI{32.5}{\micro\second} (excluding the overhead).
By defining the \textit{target} as 90\% of the best-known value (in this case, the SA result) and $P_S$ is the percentage of results reaching the target, the \textbf{time-to-target} (TTT) \cite{goto2021high} can be expressed as:
\begin{equation}
    TTT = T_{com}\frac{\log_{10}(1-0.9)}{\log_{10}(1-P_S)} = \SI{39.5}{\micro\second}\, ,
    \label{eq:time_to_target}
\end{equation}
where $T_{com}$ is the computation time.
\section{Conclusions} \label{sec:Conclusions}
The growing interest in solving CO problems has stimulated the development of efficient Ising machines. Among these, SBMs are particularly attractive due to their ability to find approximate solutions to large-scale Ising problems within a limited amount of time and their parallelizability, making them well-suited for hardware implementations.\\
This article presents an open-source hardware architecture implementing dSB algorithm and its heated version for low-tier FPGAs. To the best of our knowledge, this is the first open-source hardware implementation of the dSB algorithm. The hardware description is highly flexible, allowing users to customize the degree of parallelization according to their specific requirements. A proof-of-concept implementation capable of solving 256-variable problems was achieved on the \textbf{AMD Kria KV260} SoM FPGA and validated using well-known benchmarks such as the max-cut and knapsack problems.\\
Several challenges remain in improving the architecture’s efficiency. These include mitigating data dependencies between the matrix-vector multiplication and the time evolution steps, potentially through approximations in the motion equations to enable further parallelization. Additionally, implementing on-board Random Number Generators (RNGs) for initializing position and momentum values could significantly reduce initialization overhead. To extend the architecture’s applicability to real-world scenarios, preconditioning methods for adapting problem coefficients to the numeric representation of the architecture and optimizing algorithm parameters are also required. Such advancements would open the way for an ASIC version.\\
In conclusion, this work represents a significant step toward practical hardware implementations of quantum-inspired algorithms for combinatorial optimization. Offering an open-source, flexible solution for FPGAs opens the door to broader adoption and further innovation in hardware-accelerated optimization, with future improvements promising even greater scalability and efficiency.

%% file: pseudocode.tex
{ \centering
\begin{minipage}{0.7\linewidth}
\begin{algorithm}[H]
\caption{Simulated Bifurcation}
\label{alg:sequential}

\SetAlgoNlRelativeSize{-1}

\SetKw{KwTo}{to}
\SetKwFor{Forp}{for}{do}{}
\textbf{init} $\underline{\underline{J}}$\;
\textbf{rand} $\underline{x}$, $\underline{y}$\;
$\Delta a \leftarrow a_0/n_\textrm{steps}$\;

\For{$k=0$ \KwTo $n_\textrm{steps}-1$}{
    \For{$i=0$ \KwTo $n_\textrm{spin}-1$}{
        \For{$j=0$ \KwTo $n_\textrm{spin}-1$}{\tikzmark{startfor}
            \eIf{$dSB$}{
            $acc_i \leftarrow acc_i + J_{i,j} \cdot \text{sgn}(x_j)$\;
            }{$acc_i \leftarrow acc_i + J_{i,j} \cdot x_j$\;}
        }
        \If{HEATING}{
            $y_\textrm{TMPi} \leftarrow y_i$\;
        }
    }\tikzmark{endfor}
\begin{tikzpicture}[overlay, remember picture]
    \draw[green!40!white, thick, rounded corners, fill, nearly transparent] ([shift={(6.76cm,8.0ex)}]pic cs:startfor) rectangle ([shift={(-0.2cm,0.4cm)}]pic cs:endfor);
    \node[anchor=north east, font=\bfseries\color{green!60!black}\large] at ([shift={(7.7cm,1cm)}]pic cs:endfor) {MM};
\end{tikzpicture}

\For{$i=0$ \KwTo $n_\textrm{spin}-1$}{\tikzmark{startfor2}
    $y_i \leftarrow y_i + ((a-a_0) \cdot x_i + c_0 \cdot acc_i) \cdot \Delta_t$\;
    $x_i \leftarrow x_i + a_0 \cdot y_i \cdot \Delta_t$\;
    
    \If{$|x_i|>1$}{
        $x_i \leftarrow \text{sgn}(x_i)$\;
        $y_i \leftarrow 0$\;
    }
    
    \If{HEATING}{
        $y_i \leftarrow y_i + \gamma \cdot y_{TMPi} \cdot \Delta_t$\;
    }

}\tikzmark{endfor2} $a \leftarrow a + \Delta a$\;
} 
\begin{tikzpicture}[overlay, remember picture]
    \draw[red!40!white, thick, rounded corners, fill, nearly transparent] ([shift={(7.4cm,5.6ex)}]pic cs:startfor2) rectangle ([shift={(-.2cm,0.4cm)}]pic cs:endfor2);
    \node[anchor=north east, font=\bfseries\color{red!60!black}\large] at ([shift={(7.7cm,1cm)}]pic cs:endfor2) {TE};
\end{tikzpicture}

\end{algorithm}
\end{minipage}
\par }

%% file: alg_parallelization/alg_parallelization.tex
Three degrees of parallelization have been introduced in the hardware implementation to improve the dSB algorithm execution time. In particular, the presence of a Matrix-vector Multiplication (MM) provides opportunities for parallelization in hardware implementation. Indeed, applying the unrolling technique to MM allows the simultaneous processing of multiple rows and columns of the \textbf{J} matrix, as depicted in Fig.~\ref{fig:matrix_parallel}.  Each element ($\textrm{MM}_i$) of the result vector is obtained by computing the product between the $i^{\textrm{th}}$ row of the \textbf{J} matrix and $\mathbf{\Tilde{x}}$, which represents the vector whose $i^{\textrm{th}}$ element is the sign value of $x_i$. The expression for calculating $\textrm{MM}_i$ is reported in the following:
\begin{equation}
    \textrm{MM}_i = \sum_{j=0}^{n_\textrm{spin}-1} \textbf{J}_{ij} \cdot \sgn{(x_j)} \, .
\end{equation}
\par
The first unrolling consists of considering $\mathbf{P_c}$ elements of a \textbf{J} row and multiplying them for the respective $\mathbf{P_c}$ elements of $\mathbf{\Tilde{x}}$. The partial products ($J_{ij}\cdot \sgn(x_j)$) of the vector-vector multiplication are sent to a $\mathbf{P_c}$-operand adder capable of performing $\mathbf{P_c}$ additions in a single step. This unrolling reduces the time required for the computation of MM (Algorithm \ref{alg:sequential}) by a factor $\mathbf{P_c}$, as highlighted in Equation~\ref{eq:time_Pc}.
\begin{equation}
    t_{\textrm{exec1}} = \mathcal{O} \biggl(n_\textrm{steps}\cdot n_\textrm{spin} \cdot \left[ \frac{n_\textrm{spin}}{\mathbf{P_c}} + 1\right]\cdot T_{\textrm{ck}} \biggr) \, .
    \label{eq:time_Pc}
\end{equation}
Multiple rows of the \textbf{J} matrix can be processed in parallel leading to a second unrolling. Defining as $\mathbf{P_r}$ the amount of vector-vector multiplications performed simultaneously by $\mathbf{P_r}$ units, the time required for the execution of the algorithm is further reduced by a factor $\mathbf{P_r}$ as shown in Eq. \ref{eq:time_Pr}, as $\mathbf{P_r}$ results ($\textrm{MM}_i$) are concurrently generated.
\begin{equation}
    t_{\textrm{exec2}} = \mathcal{O} 
 \biggl(n_\textrm{steps}\cdot n_\textrm{spin} \cdot \left[\frac{n_\textrm{spin}}{\mathbf{P_c}\cdot\mathbf{P_r}} + 1\right]\cdot T_{\textrm{ck}} \biggr) \, .
    \label{eq:time_Pr}
\end{equation}
The pseudocode of dSB after the two unrollings is shown in Algorithm \ref{alg:unrolled}.

\par 
\noindent Lastly, a third degree of parallelization ($\mathbf{P_b}$) can be introduced by dividing the $\textbf{J}$ matrix into $\mathbf{P_b}$ blocks. Each of them is characterized by the two aforementioned degrees of parallelization. In particular, by defining a Matrix-vector Multiplication Time Evolution (MMTE) unit capable of performing both MM and TE operations, and replicating it $\mathbf{P_b}$ times, it is possible to improve the execution time of the algorithm by a factor $\mathbf{P_b}$, dividing both MM and TE terms of Eq. \ref{eq:sequential}. Therefore:
\begin{equation}
    t_{\textrm{exec3}} = \mathcal{O} 
 \biggl(n_\textrm{steps}\cdot \frac{n_\textrm{spin}}{\mathbf{P_b}} \cdot \left[\frac{n_\textrm{spin}}{\mathbf{P_c}\cdot\mathbf{P_r}} + 1\right]\cdot T_{\textrm{ck}} \biggr) \, .
    \label{eq:time_Pb}
\end{equation}
\input{alg_parallelization/pseudocode_unrolled}

\subsection{MM and TE overlapping}
Blocks MM and TE of Algorithm \ref{alg:sequential} can be executed in a pipeline fashion. After the second unrolling of the algorithm, $\mathbf{P_r}$ results (MM$_i$) are generated in parallel and used as inputs by the TE block to update the variables $x_i$ and $y_i$. Specifically, it is possible to overlap the time required by the TE block to update the variables $x_i$ and $y_i$ with the time required by the MM block to generate the MM$_i$ results. After unrolling, each MM block produces $\mathbf{P_r}$ values in $\frac{n_\textrm{spin}}{\mathbf{P_c}}$ clock cycles. Assuming that each $x_i,y_i$ pair is computed in one clock cycle, the TE block needs $\mathbf{P_r}$ cycles to update $\mathbf{P_r}$ pairs. Therefore, TE and MM can work in pipeline if the following expression is satisfied:
\begin{equation}
	\frac{n_\textrm{spin}}{\mathbf{P_c}} \ge \mathbf{P_r}
	\label{eq:overlap_dis}
\end{equation}
\begin{figure}[htpb]
    \centering
    \includegraphics[width=0.7\textwidth]{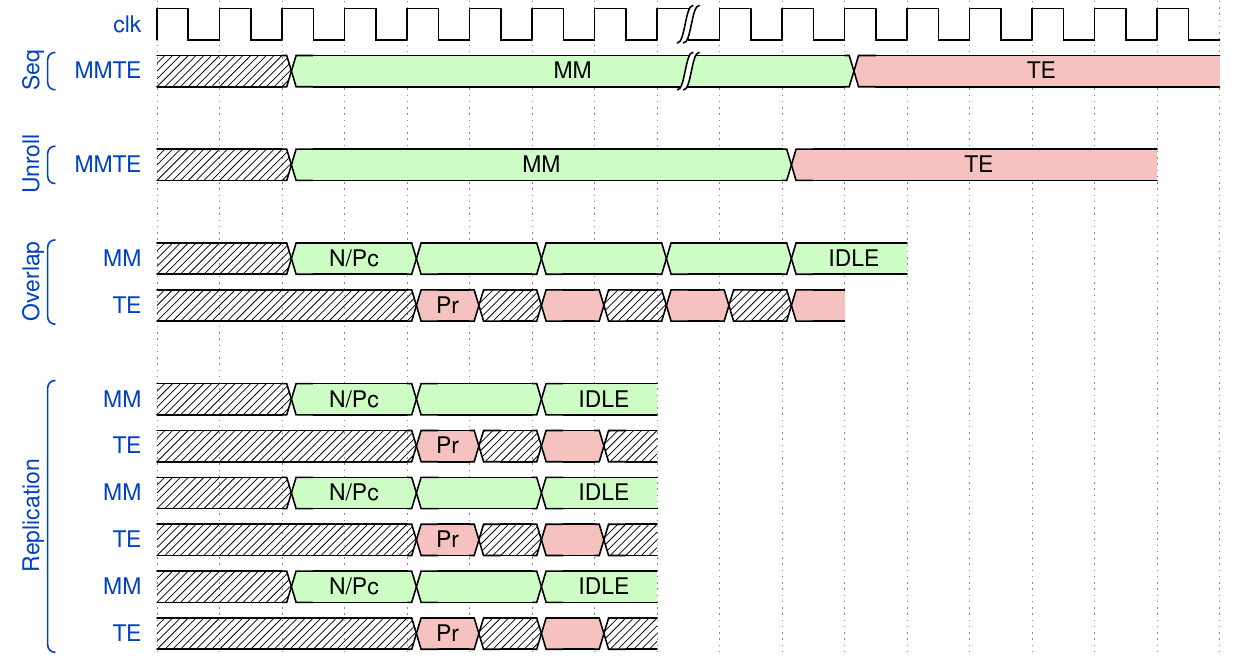}
    \caption{At the top, the timing of the sequential algorithm is illustrated. In the second timing, MM is sped up by a factor \mbox{$\mathbf{P_c}\cdot \mathbf{P_r}$} after unrolling. In the third timing, MM and TE overlapping is implemented. On the bottom, MMTE replication is also applied with $\mathbf{P_b}$ equal to three.}
    \label{fig:timing_comparison}
\end{figure}
Fig. \ref{fig:timing_comparison} illustrates the timing differences between the sequential execution (shown at the top) and the pipelined execution (third diagram) of the algorithm. In the pipelined execution, the TE block is almost entirely overlapped with the MM block, except for the initial $\frac{n_\textrm{spin}}{P_c}$ needed to fill the pipeline. Consequently, the new execution time can be expressed as:
\begin{equation}
    t_{\textrm{exec4}} = \mathcal{O} \biggl( n_\textrm{steps}\cdot \frac{n_\textrm{spin}}{\mathbf{P_b}}\cdot \left[\frac{n_\textrm{spin}}{\mathbf{P_c}\cdot\mathbf{P_r}} + \frac{1}{\mathbf{P_c}}\right]\cdot T_{\textrm{ck}} \biggr)  \, .
    \label{eq:time_pipeline}
\end{equation}

\subsection{Degrees of parallelization choice }
In Sec. \ref{sec:alg_parall}, three degrees of parallelization have been presented. Their values have been selected to minimize the algorithm execution time in accordance with the available hardware resources. From Eq. \ref{eq:overlap_dis}, the perfect overlap between MM and TE is ensured when $\frac{n_\textrm{spin}}{\mathbf{P_c}} = \mathbf{P_r}$. Table \ref{tab:parameters} presents various potential choices for the three parameters, along with the relative estimated execution times for a 256-spin Ising problem. The solution highlighted in red offers the lowest latency, while the green one provides the best trade-off between allocated resources and speed.

\par The first choice requires the fewest cycles per step but involves accessing 64 elements of $\mathbf{J}$ in parallel. This necessitates a high memory data width and 64-operand adders, which may slow down the hardware implementation due to their high delay. The second choice (in green) represents a good compromise between speed and hardware resources and has been selected for the FPGA implementation. The disadvantage in terms of clock cycles per step is small compared to the first solution. On the contrary, utilizing fewer hardware resources can result in a smaller routing delay when mapping the architecture within the FPGA, potentially leading to a higher clock frequency.
\par
Indeed, the maximum level of parallelization is reached when all the rows of the $\mathbf{J}$ matrix are processed simultaneously. For a 256-spin problem, this could correspond to $P_r = P_c = P_b = 16$. However, the required resources might exceed the available ones.

\begin{figure}[t]
\centering
\includegraphics[width=\textwidth]{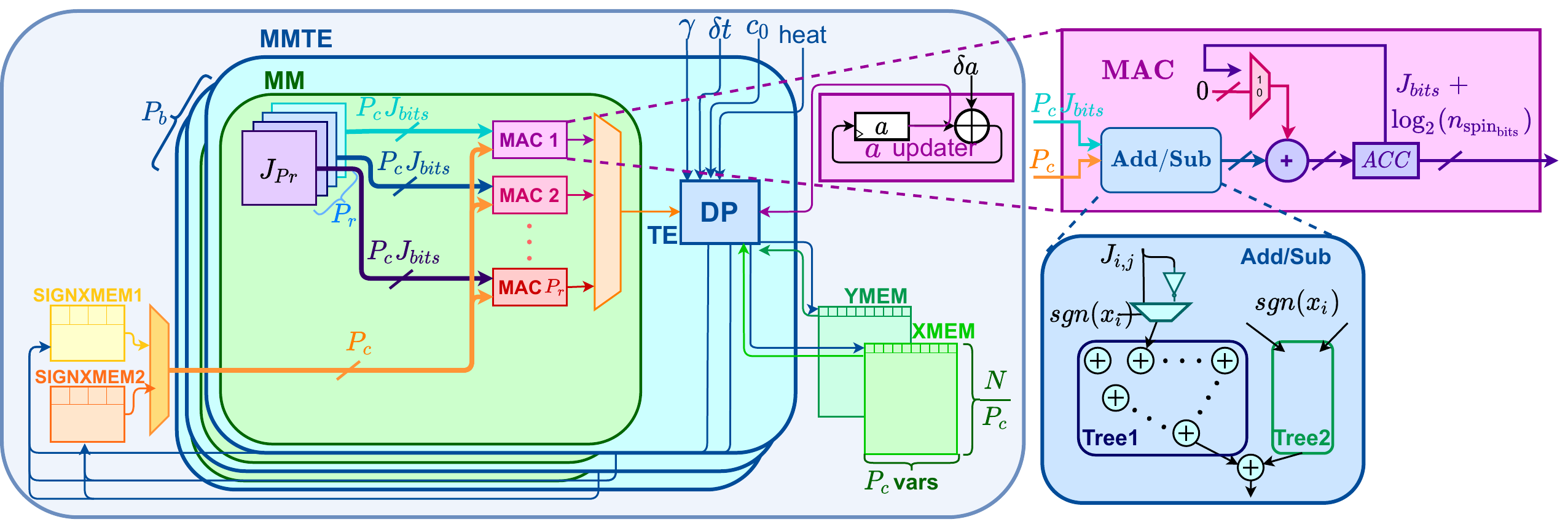}
	    \caption{High-level description of the proposed architecture. It is composed of $P_b$ Matrix-vector Multiplication Time Evolution (MMTE) blocks evaluating the oscillator state evolution. Each of them includes a Matrix-vector Multiplication (MM) block, composed of Multiply-ACcumulate (MAC) units,  allowing the parallelization of operation by a factor $P_r$. On the right, a zoom of the MAC and Adder Subtractor block are provided. }
	    \label{fig:Architecture}
\end{figure}

%% file: alg_parallelization/pseudocode_unrolled.tex
{ \centering
\begin{minipage}{0.8\linewidth}
\begin{algorithm}[H]
\caption{Algorithm after second unrolling}
\label{alg:unrolled}
\SetKw{KwTo}{to}
\SetKw{in}{in}
\SetKw{range}{range($0,n_\textrm{spin},\mathbf{P_c}$)}
\setlength{\leftskip}{0pt} 
\For{$i=0$ \KwTo $n_\textrm{spin}-1$}{\tikzmark{startfor_alg2}
    \For{
        $j$ \in \range}{
        $\textrm{mul}_0 \leftarrow J_{i,j}\cdot \sgn(x_j)$\;
        \hspace{2cm}$\vdots$\\
        $\textrm{acc}_i \leftarrow \textrm{acc}_i + \sum_{c=0}^{\mathbf{P_c}-1} \textrm{mul}_c$\;
    }
    \hspace{2cm}$\vdots$\\
    \For{
        $j$ \in \range}{
        $\textrm{mul}_0 \leftarrow J_{i+\mathbf{P_r}-1,j}\cdot \sgn(x_j)$\;
        \hspace{2cm}$\vdots$\\
        $\textrm{acc}_{i+\mathbf{P_r}-1} \leftarrow \textrm{acc}_{i+\mathbf{P_r}-1} + \sum_{c=0}^{\mathbf{P_c}-1} \textrm{mul}_c$\;
}\tikzmark{endfor_alg2}
}
\end{algorithm}
\begin{tikzpicture}[overlay, remember picture]
    \draw[green!40!white, thick, rounded corners, fill, nearly transparent] ([shift={(9.5cm,5ex)}]pic cs:startfor_alg2) rectangle ([shift={(-1.2cm,-0.6cm)}]pic cs:endfor_alg2);
    \node[anchor=north east, font=\bfseries\color{green!60!black}\large] at ([shift={(7.4cm,0.25cm)}]pic cs:endfor_alg2) {MM};
\end{tikzpicture}
\end{minipage}
\par }

%% file: alg_parallelization/matrix.tex
\begin{figure}[htpb]
\centering
    \[
    \left[\begin{array}{c}
         \textrm{MM}_0\tikzmark{matPcstarty}\\
         \vdots\\
         \textrm{MM}_{Pr-1}\tikzmark{matPcendy}\\
         \\
         \\
         \\
         \\
         \vdots\\
         \\
         \\
         \textrm{MM}_{N-1}
    \end{array}\right]
    =
    \left[\begin{array}{cccccc}
    J_{00}\tikzmark{matPcstart} & J_{01}  & \ldots & J_{0,Pc-1}\tikzmark{matPcend} & \ldots & J_{0,N-1}\\
    \vdots &  & & & & \\
    J_{Pr-1,0} & & & & & \\
     &  &  & &  &  \\
    &&&&&\\
    &&&&&\\
    &&&&&\\
    &&&&&\\
    &&&&&\\
    &&&&&\\
    J_{N-1,0} & & & & & \\
    \end{array}\right]
    \left[\begin{array}{c}
        \tilde{x}_0\tikzmark{matPcstart2}  \\
         \vdots\\
         \tilde{x}_{Pc-1}\tikzmark{matPcend2}\\
         \tikzmark{matPcstarty3}\\
         \\
         \tikzmark{matPcendy3}\\
         \\
         \vdots\\
         \\
         \\
         \tilde{x}_{N-1}
    \end{array}\right]
\]
\begin{tikzpicture}[overlay, remember picture, show background rectangle]
    
    \draw[blue!80!black!80!black, thick, rounded corners, line width=1pt] ([shift={(-.95cm,3ex)}]pic cs:matPcstart) rectangle ([shift={(2.4cm,-8ex)}]pic cs:matPcend);
    \draw[green!50!black, thick, rounded corners, line width=1pt] ([shift={(-1.05cm,3.5ex)}]pic cs:matPcstart) rectangle ([shift={(2.5cm,-13.5ex)}]pic cs:matPcend);
    \draw[green!50!black, thick, rounded corners, line width=1pt] ([shift={(-1.05cm,-14ex)}]pic cs:matPcstart) rectangle ([shift={(2.5cm,-31ex)}]pic cs:matPcend);
    \draw[green!50!black, thick, rounded corners, line width=1pt] ([shift={(-1.2cm,4ex)}]pic cs:matPcstarty) rectangle ([shift={(0.2cm,-7ex)}]pic cs:matPcendy);
    \draw[blue!80!black, thick, rounded corners, line width=1pt] ([shift={(-1.1cm,3ex)}]pic cs:matPcstarty) rectangle ([shift={(0.1cm,-2ex)}]pic cs:matPcendy);
    \draw[green!50!black, thick, rounded corners, line width=1pt] ([shift={(-1.2cm,-14ex)}]pic cs:matPcstarty) rectangle ([shift={(0.2cm,-25ex)}]pic cs:matPcendy);
    \draw[orange!80!white, thick, rounded corners, fill, nearly transparent] ([shift={(-.8cm,3ex)}]pic cs:matPcstart) rectangle ([shift={(0.2cm,-2ex)}]pic cs:matPcend);
    \draw[orange!80!white, thick, rounded corners, fill, nearly transparent] ([shift={(-.9cm,2ex)}]pic cs:matPcstart2) rectangle ([shift={(0.2cm,-1.5ex)}]pic cs:matPcend2);
    \draw[decorate, decoration={brace, amplitude=5pt, raise=6pt, aspect=0.5}, line width=1.5pt, color=orange!80!white]
        ([shift={(-.8cm,3ex)}]pic cs:matPcstart) -- node[above=10pt, color=orange!80!white] {\scriptsize Pc} ([shift={(0.2cm,3ex)}]pic cs:matPcend);
    \draw[decorate, decoration={brace, amplitude=5pt, mirror, raise=8pt, aspect=0.5}, line width=1.5pt, color=blue!80!black]
        ([shift={(-1.3cm,3ex)}]pic cs:matPcstarty) -- node[left=12pt, color=blue!80!black] {\scriptsize Pr} ([shift={(-1.6cm,-2ex)}]pic cs:matPcendy);
    \draw[decorate, decoration={brace, amplitude=5pt, mirror, raise=28pt, aspect=0.5}, line width=1.5pt, color=green!50!black]
        ([shift={(-1.1cm,3ex)}]pic cs:matPcstarty) -- node[left=34pt, color=green!50!black] {\scriptsize Pb} ([shift={(-1.4cm,-25ex)}]pic cs:matPcendy);
\end{tikzpicture}
\caption{Matrix-vector multiplication between the \textbf{J} matrix and the $\mathbf{\tilde{x}}$ vector. The three degrees of parallelization $Pc$, $Pr$ and $Pb$ are highlighted. $\tilde{x_i}$ indicates the sign value of the $x_i$ variable.}
\label{fig:matrix_parallel}
\end{figure}